\documentstyle[floats,aps,epsf,prl]{revtex}
\epsfclipon

\begin{document}
\draft
\twocolumn[\hsize\textwidth\columnwidth\hsize\csname @twocolumnfalse\endcsname

\title{Spatial and Spectral Multifractality of the Local Density of
  States at the Mobility Edge} 

\author{Bodo Huckestein and Rochus Klesse}

\address{Institut f\"ur Theoretische Physik, Universit\"at zu
  K\"oln, D-50937 K\"oln, Germany}  
\date{\today}

\maketitle

\begin{abstract}
  
  We performed numerical calculations of the local density of states
  (LDOS) at disorder induced localization-delocalization
  transitions. The LDOS defines a spatial measure for fixed energy and
  a spectral measure for fixed position.  At the mobility edge both
  measures are multifractal and their generalized dimensions $D(q)$ and
  $\tilde{D}(q)$ are found to be proportional: $D(q)=d\tilde{D}(q)$,
  where $d$ is the dimension of the system. This observation is
  consistent with the identification of the frequency-dependent length
  scale $L_\omega \propto \omega^{-1/d}$ as an effective system size.
  The calculations are performed for two- and three-dimensional
  dynamical network models with local time evolution operators. The
  energy dependence of the LDOS is obtained from the time evolution of
  the local wavefunction amplitude of a wave packet, providing a
  numerically efficient way to obtain information about the
  multifractal exponents of the system.

\end{abstract}
\vskip2pc

]

In the presence of disorder electronic states in phase-coherent
systems can get localized. For dimensions $d\geq2$ such a 
disordered system can exhibit a transition from localized to extended
eigenstates as a function of energy or strength of the disorder,
similar to continuous thermodynamic phase transitions
\cite{LR85KM93}. This transition is accompanied by a diverging length scale
$\xi(E)\propto |E-E_c|^{-\nu}$ that can be identified for localized
states with the localization length. Here $E_c$ is the critical energy
of the transition. On length scales shorter than $\xi(E)$ the
eigenstates exhibit strong fluctuations \cite{Weg80}. Sufficiently
close to the transition the localization length exceeds the system
size so that the wave functions fluctuate on all length scales up to
the system size $L$. The squared modulus of these critical eigenstates
forms a multifractal measure $\rho({\bf r})=|\psi({\bf r})|^2$
\cite{CP86-Jan94}. The scaling behavior of
multifractal measures is determined by an infinite set of exponents
$D(q)$, called generalized dimensions. The multifractal fluctuations
lead to anomalous behavior of the diffusion coefficient
\cite{CD88,HS94}. 

The localization properties of the wavefunction are also reflected in the
spectral properties of the system. Localized states correspond to a
pure point like spectrum, extended states to an absolutely continuous
spectrum, and critical states to a singular continuous spectrum. The
latter kind of spectrum was observed in quasiperiodic systems
\cite{AA80-MDA96}, systems on incommensurate structures
\cite{Sok85}, in crystals in a magnetic field \cite{Azb64Hof76}, and
also in random one-dimensional systems \cite{RJLS95JL96}. For these
systems the spectral measure excited by a wave packet is multifractal.
Of particular interest has been the implications of the local spectra
for dynamical properties \cite{KPG92,Gua93GM94}. Ketzmerick,
Petschel, and Geisel \cite{KPG92} found that the return probability of
a wave packet decays in time $t$ asymptotically as $t^{-\tilde{D}(2)}$,
where $\tilde{D}(2)$ is the correlation dimension of the associated
spectral measure, the local density of states (LDOS).

The two aspects of spatial and spectral multifractality were brought
together by Huckestein and Schweitzer who showed that enhanced return
probability of wavepackets at the mobility edge of quantum Hall
systems could be interpreted both from the spectral \cite{KPG92} as
well as the spatial properties of the local density of states
\cite{HS94}. They showed that the generalized dimensions $D(2)$ and
$\tilde{D}(2)$ characterizing the second moments of the spatial and
spectral measures, respectively, are related by the spatial dimension
$d=2$ of the system, $D(2)=d\tilde{D}(2)$. This follows from dynamical
scaling by which the two-particle spectral function $S(q,\omega)$
becomes a function of $qL_\omega$ at criticality \cite{CD88}. Here,
$L_\omega=(\rho(E_c)\omega)^{-1/d}$ can be interpreted as the system
size with mean level spacing $\omega$ and $\rho(E_c)$ is the density
of states at the mobility edge \cite{otherlength}. The length scale
$L_\omega$ introduced by the finite frequency $\omega$ cuts off
correlations and acts as an effective system size \cite{Zir94FPJ96}.
Recently, the same relation was shown to hold in three dimensional
systems at the Anderson transition\cite{BHS96}.

In this letter, we show that at the mobility edge the spatial and
spectral structures of the LDOS are intimately related: their
respective generalized dimensions $D(q)$ and $\tilde{D}(q)$ are
proportional, \begin{equation}
  \label{result}
  D(q)=d\tilde{D}(q),
\end{equation}
where $d$ is the space dimension of the system. This relation,
generalizing the result of ref.~\cite{HS94} to arbitrary $q$, is a
consequence of the length scale $L_\omega$ acting quite generally as
an effective system size, not only for correlation functions but also
for moments of the LDOS. We test the relation numerically for two- and
three-dimensional network models \cite{CC88,CD95}. Introducing a time
evolution into these models \cite{KM95} allows us to calculate the
LDOS by studying the time evolution of wave packets. This provides a
new efficient method to obtain the multifractal exponents of the LDOS
without the need to diagonalize large matrices.

Before motivating the relation (\ref{result}) let us define our
multifractal analysis. We study a normalized multifractal density
$\rho(x)$ on a $d$-dimensional domain $\Omega$. Here, $\Omega$ is the
two- or three-dimensional space or the one-dimensional energy axis,
depending on whether we study the spatial or spectral aspect of the
LDOS. We study the scaling of the box probabilities
\begin{equation} %nn
P_i(l) = \int_{\Omega_i(l)} d^dx \rho(x)
\end{equation} %nn
with respect to the box sizes $l$. $\Omega_i(l)$ is the volume of
the $i$th of $N_l= (L/l)^d$ non-overlapping hypercubes of linear size
$l$ covering the whole domain $\Omega$ of linear size $L$.
For a multifractal density the averaged $q$th moments,
$
\langle P^q(l,L)\rangle = N^{-1}_l \sum_{i=1}^{N_l} \left(P_i(l)\right)^q,
$
show power law dependence on $l/L \ll 1$ for all real $q$,
\begin{equation} %nn
\langle P^q(l,L)\rangle \propto \left(\frac{l}{L}\right)^{d+\tau(q)}.
\end{equation} %nn
The distinguishing feature of a multifractal is that the exponents
$\tau(q)\equiv(q-1)D(q)$ are a non-linear function of $q$. A multifractal
is thus described by a non-countable set of exponents $D(q)$. If the
multifractal measure is taken from a statistical ensemble, the system
average over the box probabilities can be replaced by an ensemble
average for a single box.

The LDOS is given by $\rho({\bf r},E)=\sum_{\alpha}
\delta(E-E_\alpha)|\phi_\alpha({\bf r})|^2$, where the sum runs
over the quantum numbers $\alpha$ with corresponding eigenstates
$\phi_\alpha$.  At energies $E$ close to the mobility edge $E_c$ the
LDOS exhibits multifractal properties on length scales between
microscopic lengths, like the lattice constant $a$ or the elastic mean
free path, and the system size $L$ or the correlation length $\xi(E)$.
The corresponding lower and upper energy scales are the mean level
spacing $\Delta = \rho(E_c)^{-1}L^{-d}$ and $E_\xi\propto L^{-1/\nu}$,
respectively. For energies $\omega$ less than $E_\xi$ the correlation
length $\xi(\omega)$ exceeds the system size $L$.

We now argue that the existence of a single frequency-dependent length
scale $L_\omega$ implies the relation (\ref{result}). Consider box
probabilities of the LDOS with respect to both space and energy
\begin{equation}
  \label{ldos_box}
  P_i(l,\omega) = \int_{\Omega_i(l)}d^dx
  \int_{E_c-\frac{\omega}{2}}^{E_c+\frac{\omega}{2}} dE\,
  \rho\left({\bf r},E\right),
\end{equation}
where we consider an energy interval in the critical region described
above. For energies $\omega$ less than the mean level spacing $\Delta$
these box probabilities scale like single eigenfunctions, $\langle
P^q(l,\omega)\rangle \propto (l/L)^{d+\tau(q)}$. For energies larger
than the mean level spacing, we expect the disorder average of
the moments of these box probabilities to scale with the same
exponents but with the system size $L$ replaced by a
frequency-dependent effective system size $L_\omega$,
\begin{equation}
  \label{ldos_scale}
  \langle P_i^q(l,\omega) \rangle \propto
  \left(\frac{l}{L_\omega}\right)^{d+\tau(q)}.
\end{equation}
In principle, the frequency-dependent length scale introduced in
eq.~(\ref{ldos_scale}) could depend on $q$. However, if we assume that
there exists only one such length scale, then we can identify
$L_\omega$ with $(\rho(E_c)\omega)^{-1/d}$ by considering the first
moment
\begin{equation}
  \label{norm}
  \langle P_i(l,\omega)\rangle = \rho(E_c)\omega l^d\propto
  \left(\frac{l}{L_\omega}\right)^d.
\end{equation}
The scaling of the spectral measure defined by the LDOS follows from
eq.~(\ref{ldos_scale}) by choosing the spatial box size $l$
of the order of the microscopic length scale $\ell$, below which the
wavefunctions become smooth functions of coordinate. The scaling of
the spectral measure with respect to energy defines the exponents
$\tilde{\tau}(q)\equiv(q-1)\tilde{D}(q)$,
\begin{equation}
  \label{tilde_scale}
  \langle P_i^q(\ell,\omega) \rangle \propto
  \left(\rho(E_c)\ell^d\omega\right)^{1+\tau(q)/d}
  \propto \omega^{1+\tilde{\tau}(q)},
\end{equation}
leading to eq.~(\ref{result}).

\begin{figure}
  \begin{center}
    \leavevmode
    \epsfysize=5cm
    \epsffile[140 500 500 714]{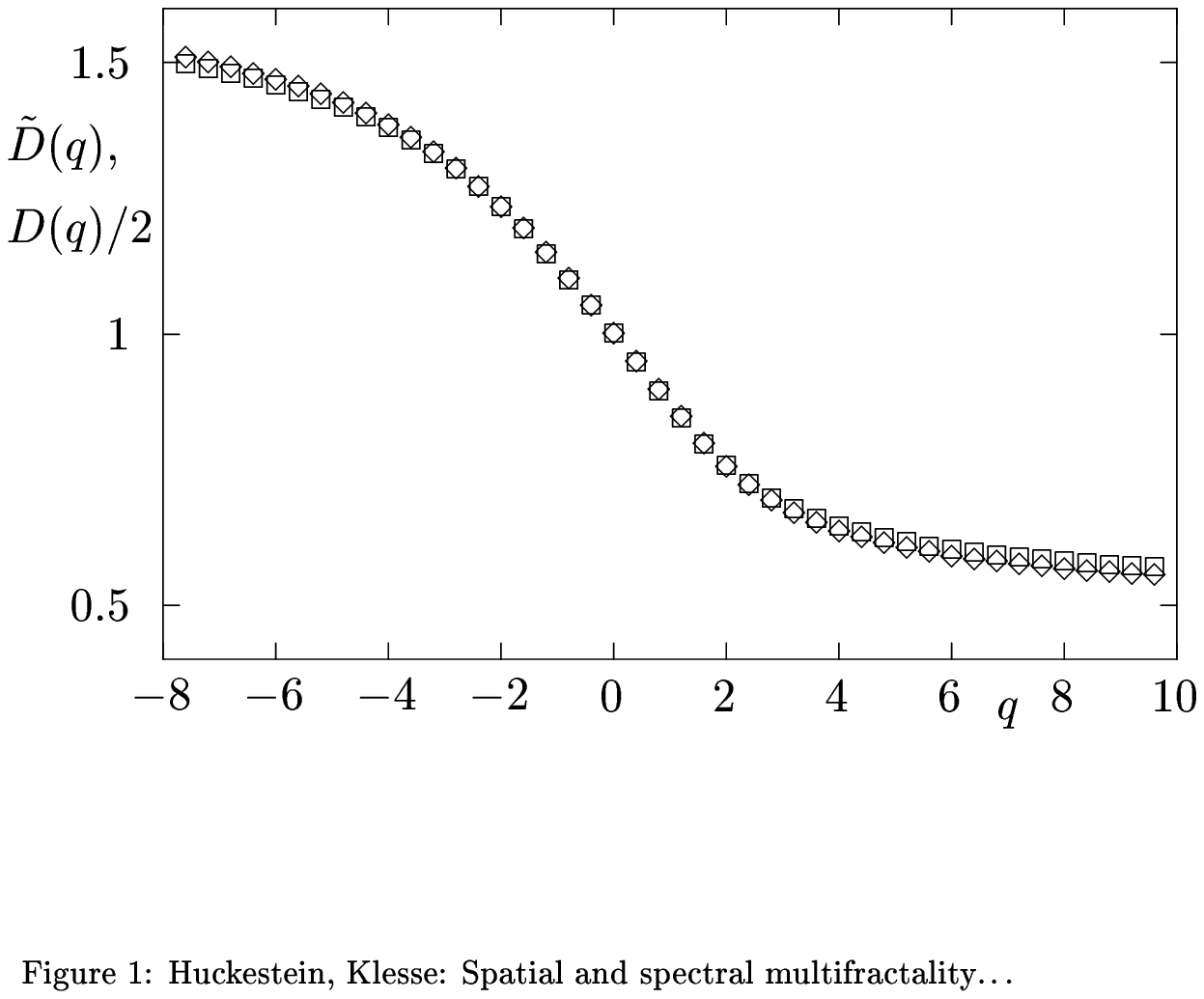}
  \end{center}
  \caption{Generalized dimensions $D(q)/2$ ($\Diamond$) and $\tilde{D}(q)$
    ($\Box$) for a two-dimensional network model at the quantum
    Hall critical point. Systems of size $150\times150$ ($\Diamond$)
    and $200\times200$ ($\Box$) were used and the statistical
    uncertainties are smaller than the symbol sizes.}
  \label{fig:2d_dq}
\end{figure}

\begin{figure}
  \begin{center}
    \leavevmode
    \epsfysize=5cm
    \epsffile[140 500 500 714]{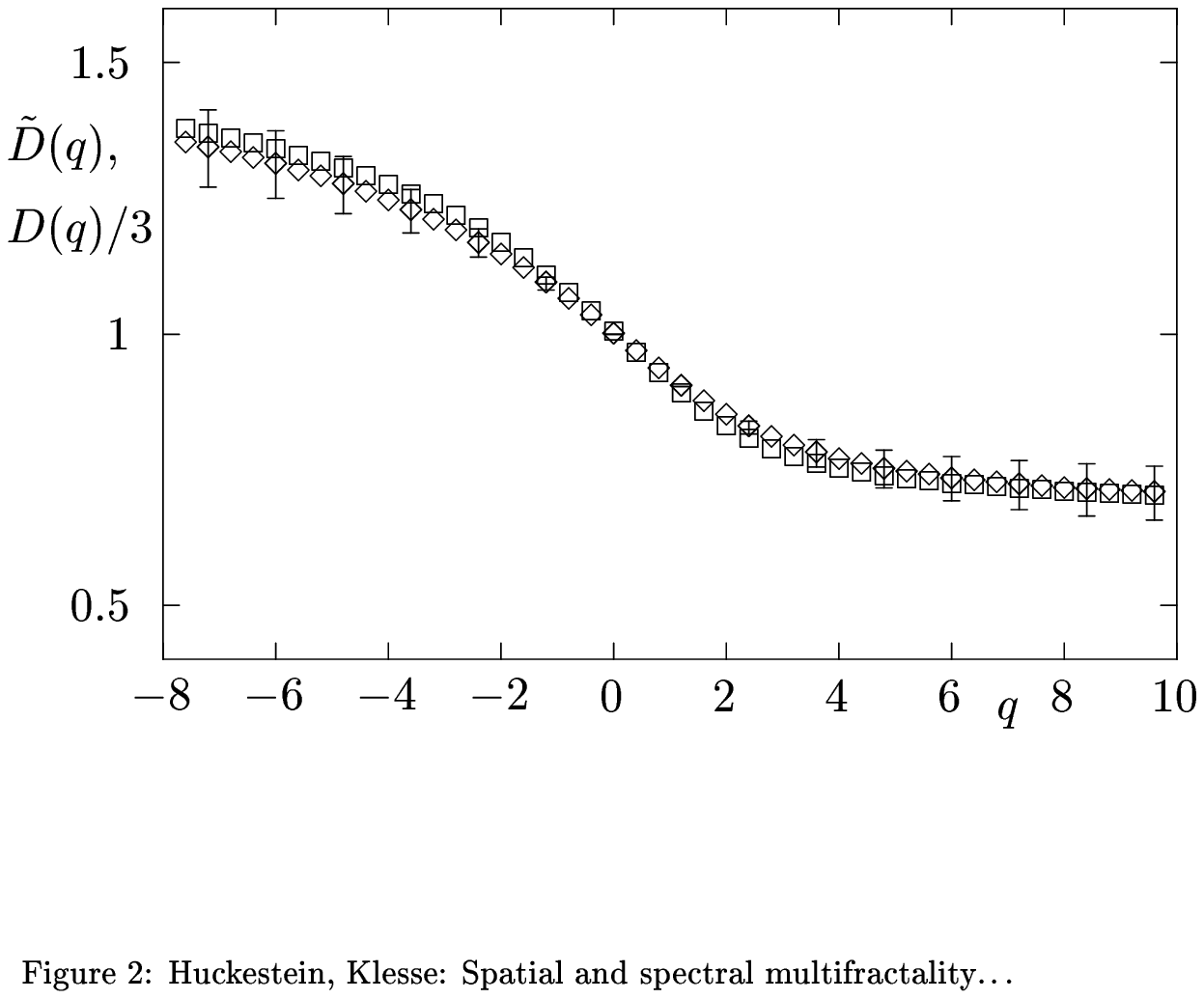}
  \end{center}
  \caption{Generalized dimensions $D(q)/3$ ($\Diamond$) and $\tilde{D}_q$
    ($\Box$) for a three-dimensional network model at the Anderson
    transition in a magnetic field. Systems sizes of $20^3$
    ($\Diamond$) and $35^3$ ($\Box$) were used and the statistical
    uncertainties in the $\tilde{D}(q)$ are less than those in the
    $D(q)/3$.}
  \label{fig:3d_dq}
\end{figure}

For our computations, we consider two normalized measures, the spatial
measure $\rho\left({\bf r}\right)=|\psi_\alpha({\bf r})|^2$ and the
spectral measure $\tilde{\rho}(E)=A^{-1}\rho\left({\bf r}_0,E\right)$.
The spatial measure is normalized since the eigenfunctions are
normalized, and the spectral measure is explicitly normalized on an
energy interval $E$, satisfying $E\ll E_\xi$. To support the validity
of eq.~(\ref{result}) we now present results of numerical
calculations. Fig.~(\ref{fig:2d_dq}) shows the functions $D(q)$ and
$2\tilde{D}(q)$ calculated for a two-dimensional network model at the
quantum Hall critical point \cite{CC88,KM95}. The functions agree
within the error bars. In order to check the dependence on the spatial
dimension $d$ of the system we study a three-dimensional network
introduced recently \cite{CD95}. In contrast to the two-dimensional
network here a band of extended states appears. At the mobility edge
of this system we again find good agreement with eq.~(\ref{result}) as
seen in Fig.~(\ref{fig:3d_dq}).

We now describe our new numerical method of obtaining the LDOS. We use
two- and three-dimensional network models, describing the integer
quantum Hall effect \cite{CC88} and the so called quantum
Hall-insulator \cite{CD95}, respectively.  Both models are extended to
{\em dynamical} network models by providing them with a unitary time
evolution operator $U$ for discrete microscopic time steps
\cite{KM95}. Within these models the evolution---involving full
quantum interference---of arbitrary initial states can be obtained
easily by iterative application of $U$. In particular, the time
evolution of a state $\psi$ initially sharply peaked at coordinate
${\bf r}_0$ yields the temporal Green's function $G({\bf r}_0,{\bf
  r}_0;t)$. The LDOS is then the Fourier transform of this quantity.
This provides an efficient method of calculating the LDOS, without the
need to diagonalize the operator $U$ or the associated Hamiltonian.

Starting point of our calculations is the 2D-network model introduced
by Chalker and Coddington \cite{CC88} to describe non-interacting
spinless electrons in the integer Quantum Hall regime. It exhibits a
localization-delocalization-transition characteristic for the QHE:
except for the critical energy $E_c$ all states are localized, with a
localization length $\xi\propto |E-E_c|^{-\nu}$, $\nu = 2.3$. Origin
of the model and exact definitions can be found in refs. \cite{CC88}
and \cite{KM95}, here we give only a brief description restricted to
our purposes.

The network consists of $2\times 2$ scattering matrices as nodes which
are arranged on a square lattice and connected by one-dimensional
unidirectional channels, called links. The scattering matrices contain
coefficients $t_{lm}$ describing transitions from electron states on
incoming links $\psi_l$ to outgoing link states $\psi_m$.  At the
critical point the transmission amplitudes are of constant value
$T_{ml}=|t_{ml}|^2= 1/2$, while the disorder is given by randomly
distributed arguments of the coefficients $t_{ml}$. {\em States} (or
{\em wave functions}) $\psi$ on the network are $N$-dimensional
complex vectors $\psi=\{\psi_l\}_{l=1,2,\ldots,N}$, where the $l$-th
component denotes the complex amplitude on link $l$. The network
operator $U$ is defined by its action on single link states $e_l=\{
\delta_{lk} \}_{k=1,\ldots,N}$,
\begin{equation} %nn
U e_l= t_{ml} e_m + t_{nl} e_n,
\end{equation} %nn
where $t_{ml}$ and $t_{nl}$ are the transmission coefficients from an
incoming link $l$ into two outgoing links $m$ and $n$ at a
node\cite{KM95}.

Here we use $U$ as the time evolution operator in discrete
time steps for states on the network. To motivate this, consider a particle
at critical energy $E_c$ in the incoming link $l$ of node $j$,
that is described by the state $\psi(0) = e_l$. After a characteristic
time $\tau$ (we choose $\tau=1$ in the following) the incident wavepacket has passed the scatterer and thereby
split into two outgoing packets in the channels $m$ and $n$ (see
fig.~1 in ref.~\cite{KM95}). This process corresponds just to the
acting of $U$ on $\psi(0)=e_l$,
\begin{eqnarray}
\label{time_step}
\psi(0)\quad \longrightarrow \quad \psi(1) &=& U\psi(0) \nonumber \\
&=& t_{ml} e_m + t_{nl} e_n. \nonumber
\end{eqnarray}
Since this happens in the same way at all scatterers we generalize relation
(\ref{time_step}) to the time evolution of an arbitrary 
state and define 
\begin{equation}
\label{evolution}
\psi(t + n) = U^n \psi(t),
\end{equation}
for integer $n$. With the latter definition the original static
network becomes a quantum-dynamical model of a disordered system. We
can relate the eigenvectors $\{ \Phi^\omega\}$ with eigenvalues $\{
e^{-i\omega}\}$ of $U$ to eigenfunctions and energy spectrum of
a Hamiltonian $H$ by identifying \cite{HC96,future_publication}
\begin{equation} %on
U=\exp(- iH\tau).
\end{equation} %on

\begin{figure}
  \begin{center}
    \leavevmode
    \epsfysize=5cm
    \epsffile[140 500 500 714]{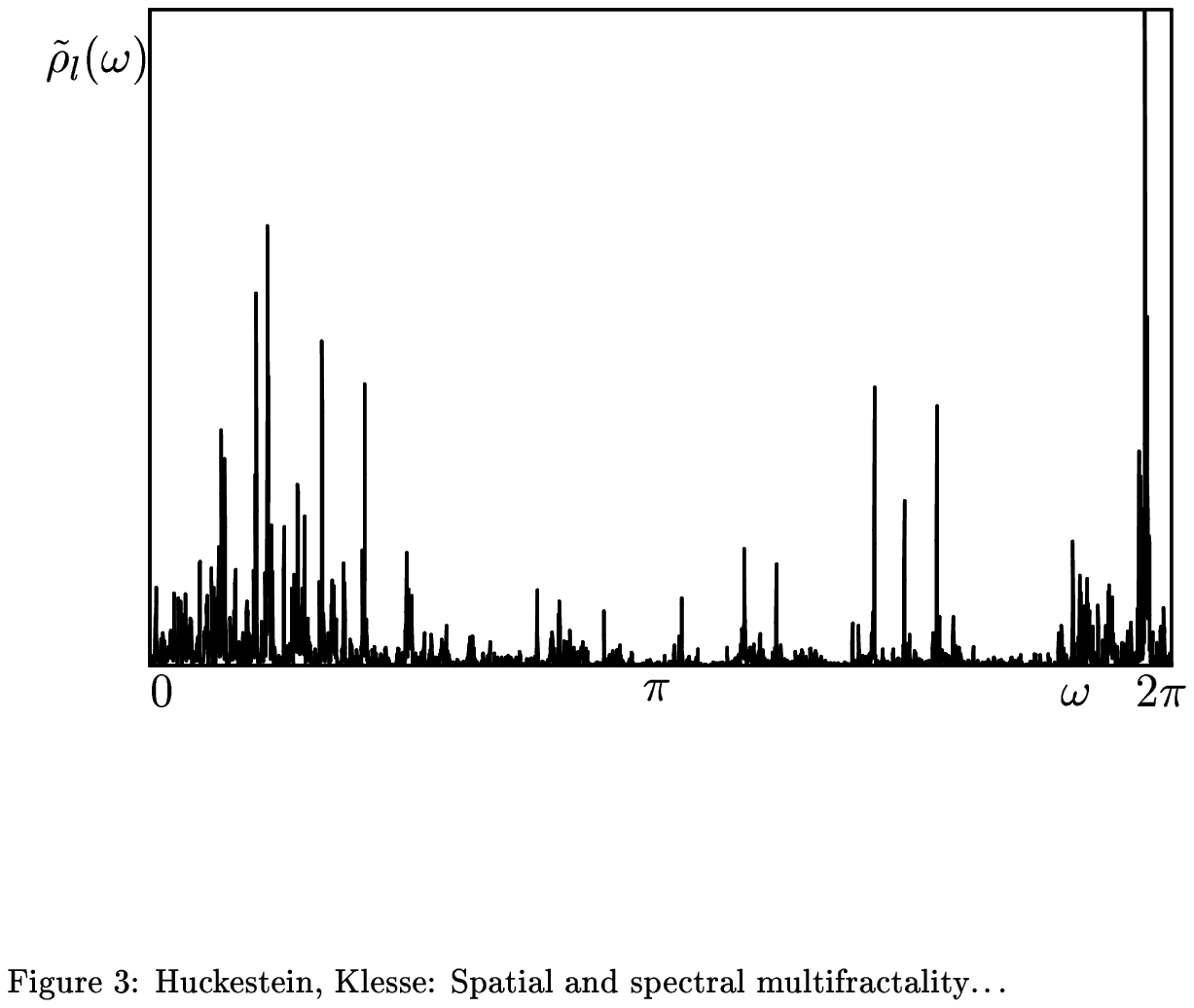}
  \end{center}
  \caption{Local density of states (LDOS) $\tilde{\rho}_l(\omega)$
    obtained from eq.~(\ref{trho}) for a quantum Hall system of
    $200\times200$ nodes. Due to the finite time $T$ (16384
    iterations) of the Fourier transform each $\delta$-function in the
    LDOS is broadened into a peak of width $\Delta\omega=2\pi/T$.  }
  \label{fig:ldos}
\end{figure}

To obtain the LDOS
$\tilde{\rho}_l(\omega)=\sum_{\omega'}\delta(\omega-\omega')
|\phi_l^{\omega'}|^2$ (the sum runs over all eigenvalues $\omega'$ of
$H$) at link $l$, we calculate numerically the temporal Green's
function $G(l,l;t)=\langle e_l|U^t e_l\rangle$. In terms of the
eigenstates of $U$ the Green's function is given by
\begin{equation} %nn 
G(l,l;t)=\sum_{\omega}  \left| \langle e_l| \phi^{\omega} \rangle
\right|^2 e^{-i\omega t}
= \int d\omega \;\tilde{\rho}_l(\omega) e^{-i\omega t}.
\end{equation} %nn 
Hence the LDOS $\tilde{\rho}_l(\omega)$ can be obtained by an inverse
Fourier transform of $G(l,l;t)$, which on  
the other hand can be generated iteratively according to (\ref{evolution}),
\begin{equation} %on
\tilde{\rho}_l(\omega) = {1 \over 2\pi} \int dt\, G(l,l;t) e^{i\omega t}
= {1\over \pi} \text{Re} \int_0^\infty dt\, G(l,l;t) e^{i\omega t}.
\label{trho}
\end{equation} %on
Fig.~(\ref{fig:ldos}) shows as an example the energy dependence of the
LDOS obtained in that way. 

The three-dimensional network investigated here is very similar to the
one studied by Chalker and Dohmen \cite{CD95}. It is built out of
layered two-dimensional networks with additional inter-layer
couplings, that we choose slightly different from Chalker and
Dohmen\cite{future_publication}. 

In conclusion, we studied numerically the spatial and spectral
multifractal measures defined by the local density of states (LDOS) at the
mobility edge of two- and three-dimensional disordered electron
systems. We have presented evidence that both of these measures are
equivalent, as the ratio of their respective generalized dimensions
$D(q)$ and $\tilde{D}(q)$ is simply given by the dimension of the
system. This result is interpreted as a consequence of the occurrence
of a single energy-dependent length scale
$L_\omega=(\rho(E_c)\omega)^{-1/d}$ acting as the effective system
size. The numerical calculations were performed for two- and three-dimensional
network models. These model, endowed with a discrete time evolution,
turned out to be especially suitable for determining the LDOS as a
function of energy. 

We gratefully acknowledge illuminating discussions with J. Hajdu,
M. Janssen, M. Metzler, and L. Schweitzer.  This work was performed
within the research program of the Sonderforschungsbereich 341 of the
Deutsche Forschungsgemeinschaft.


\begin{thebibliography}{10}
\bibitem{otherlength}
Besides $L_\omega$ a second length scale $\tilde{L}_\omega$ can be defined as
  the length that the particle diffuses in the time $1/\omega$,
  $\tilde{L}_\omega=(D/\omega)^{1/2}$. However, at the mobility edge both
  lengths are proportional to each other \cite{Cha90}.
\end{thebibliography}
\end{document}